\documentclass{article}

\usepackage{charter,lipsum,fancyvrb,ragged2e}

\usepackage{amsfonts}
\usepackage{amsmath}
\usepackage{pdflscape}
\usepackage{afterpage}

\usepackage{mathrsfs}
\usepackage{multirow}
\usepackage[table]{xcolor}
\usepackage{setspace}
\usepackage{graphicx}
\usepackage[utf8x]{inputenc}
\usepackage{newunicodechar}
\usepackage{hyperref}
\usepackage{color}
\usepackage{caption}
\usepackage[mathscr]{euscript}
\usepackage{amsthm, amssymb, amsfonts}
\usepackage{txfonts}
\usepackage{soul}
\usepackage{stmaryrd}
\usepackage{setspace}
\usepackage{txfonts}
\usepackage[ruled]{algorithm2e}
\usepackage{tabto}
\usepackage{algpseudocode}
\usepackage{pifont}

\usepackage{tcolorbox}
\usepackage{comment}
\usepackage{arydshln}

\usepackage[english]{babel}
\usepackage{multirow}
\usepackage{imakeidx}
\makeindex
\usepackage{array}

\usepackage{subfig}
\usepackage{wasysym}
\usepackage{authblk}
\usepackage{tikz}

\theoremstyle{definition}

\newcolumntype{M}[1]{>{\centering\arraybackslash}m{#1}}
\newcolumntype{N}{@{}m{0pt}@{}}

\DeclareSymbolFont{rsfs}{U}{rsfs}{m}{n}
\DeclareSymbolFontAlphabet{\mathscrsfs}{rsfs}

\RequirePackage{setspace}

\theoremstyle{definition}

\definecolor{antiquewhite}{rgb}{0.98, 0.92, 0.84}
\definecolor{babyblueeyes}{rgb}{0.94, 0.97, 1.0}

\newcounter{exam}
   
   \newcommand{\specialcell}[2][c]{%
  \begin{tabular}[#1]{@{}c@{}}#2\end{tabular}}

  \newcommand{\specials}[2][l]{%
  \begin{tabular}[#1]{@{}l@{}}#2\end{tabular}}

\addto{\captionsenglish}{%
}

\begin{document}

\title{Privacy Interpretation of Behavioural-based Anomaly Detection Approaches}

\author[$\dagger$]{Muhammad Imran Khan} 
\author[$\ddag$]{Simon N. Foley}
\author[$\dagger$]{Barry O'Sullivan}

\affil[$\dagger$]{Insight Centre for Data Analytics, School of Computer Science and Information Technology, University College Cork, Ireland.}
\affil[$\ddag$]{Department of Information Security and Communication Technology, Norwegian University of Science and Technology, Gj\o{}vik, Norway.}

\date{}

\maketitle

\begin{abstract}
This paper proposes the notion of ‘\textit{Privacy-Anomaly Detection}’ and considers the question of whether behavioural-based anomaly detection approaches can have a privacy semantic interpretation and whether the detected anomalies can be related to the conventional (formal) definitions of privacy semantics such as \textit{k}-anonymity.
The idea is to learn user's past querying behaviour in terms of privacy and then identifying deviations from past behaviour in order to detect privacy violations.
Privacy attacks, violations of formal privacy definition, based on a sequence of SQL queries (query correlations) are also considered in the paper and it is shown that interactive querying settings are vulnerable to privacy attacks based on query sequences. 
Investigation on whether these types of privacy attacks can potentially manifest themselves as anomalies, specifically as privacy-anomalies was carried out. 
It is shown that in this paper that behavioural-based anomaly detection approaches have the potential to detect privacy attacks based on query sequences (violation of formal privacy definition) as privacy-anomalies.
\end{abstract}


\section{Introduction} \label{p:1}

The recent past has witnessed an exponential increase in the volume of data being collected by organizations.
This has been enabled by the aggressive development of computing technologies.
Data is fuelling most of the revolutionary technologies.
Technological advances and widely available data have nurtured the area of data analytics.
Data analytics aims to discover meaningful insights from the data that may lead to improved decision-making. 
Data analytics offers a broad spectrum of benefits, for example, it enables a contemporary organization to anticipate business opportunities as well as enable the delivery of relevant products to its customers.
In a nutshell, analytics over a large volume of data has the potential to impact businesses and our society.

Data comes from multiple sources and may include sensitive personal data.
On the one hand, one cannot deny the importance and value of data, while on the other hand, the usage, storage, and access to this data can raise privacy concerns. 
The recently enacted EU's General Data Protection Regulation (GDPR)~\cite{53} makes it more challenging to use personal data for analytics. 
However, once the data is anonymized, it is considered to no longer be personal data~\cite{53,216}. 
Achieving anonymization is non-trivial while preserving the utility of data. 
Increasing the level of anonymization protects data but reduces utility.
Thus organizations must trade-off the need for more in-depth analytics against the privacy of individual's data.
In essence, it is a long-standing open problem to get high-quality analytics by querying the databases consisting of information about individuals while preserving the privacy of those individuals.

Numerous incidents have been reported where privacy was compromised due to poor anonymization of released data, for example, the famous case of Netflix~\cite{238}, 
AOL~\cite{237}, de-anonymization of NYC taxi data~\cite{236}, and the famous case of the Massachusetts Governor~\cite{160}.
In~\cite{160}, it was shown that by linking on shared attributes (zipcode, birth date, and gender) in two datasets, Massachusetts Group Insurance Commission's released data (considered anonymous) and voter rolls, records belonging to the Massachusetts Governor were identified.
Researchers have devised formal privacy definitions\footnote{Privacy definitions, in literature, are otherwise known as privacy models, privacy criteria, privacy metrics, privacy constraints as well as privacy principles.}~\cite{21,59,60} when these definitions are followed then the anonymized data manifests some formal guarantees.
There are several privacy definitions to anonymize data, including, \textit{k}-anonymity, \textit{l}-diversity, \textit{t}-closeness, and differential privacy.
The majority of the syntactic privacy definitions were designed for a one-time release of data.  
In contrast to these definitions, differential privacy is for interactively querying a database. 
However, differential privacy has practical limitations as well; for instance, differential privacy allows only a limited number of queries to be answered.
Allowing an unlimited number of queries results in higher noise; thus, the ability to observe correlations between attributes are lost, which is not desirable for richer analytics~\cite{186}.
Approaches to detect privacy violations, while allowing an unlimited number of queries while having richer analytical utility are desirable.

Existing work in literature on detecting malicious access (security attacks) to Database Management System (DBMS) have shown the effectiveness of behaviour-based anomaly detection system to detect these security attacks.
This work looks at to what extent these kinds of techniques can be used to detect privacy attacks. 
This work considers the research question that whether one can provide a privacy semantics for behavioural-based approaches or relate the notion of privacy-anomaly detection to the conventional definitions of privacy semantics? 
In order to answer the above-mentioned questions, the notion of  \textit{`Privacy-Anomaly Detection'}(\textit{PAD}) is introduced in this paper. 
PAD learns privacy criteria from past interactions (audit logs) and uses this criteria to check whether the current behaviour is different from past behaviour with respect to privacy. 
The PAD architecture falls within an interactive query system setting for microdata release.

We describe a na\"{i}ve instantiation of PAD using \textit{k}-anonymity privacy criteria which we refer to as (\textit{k}-anonymity)-PAD.
A study is carried out to investigate whether a security-anomaly detection system, in particular, the n-gram approach presented in~\cite{10}, can detect these (\textit{k}-anonymity)-PAD privacy-anomalies.

In this work, we also show that PAD-based interactive mechanisms are vulnerable to privacy attacks based on SQL query sequences. 
We further investigate: whether these types of privacy attacks based can potentially manifest themselves as anomalies and whether one can interpret a security-anomaly detection system in such a way that it can detect privacy attacks as privacy-anomalies.
We present the result that privacy attacks (like inferences) can be detected by applying security-anomaly detection system over the logs of interactive querying mechanisms on the basis of a PAD interpretation.

The rest of this paper is organized as follows.
Section~\ref{pad} presents a design of a privacy-anomaly detection system and 
an instantiation based on \textit{k}-anonymity. 
Section~\ref{pva} investigates whether there is a correlation between privacy and security anomalies.
Section~\ref{bpc} considers a privacy attack based on query sequence on PAD. 
Section~\ref{7.7} presents an application of security-anomaly detection system to detect (unknown) privacy attacks as privacy-anomalies.
Section~\ref{p:c} concludes this paper.

\section{Privacy-Anomaly Detection (PAD) System } \label{pad}~\index{Privacy-anomaly detection system}
This section introduce the notion of privacy-anomaly detection and present a na\"{i}ve instantiation of it based on \textit{k}-anonymity. 
We argue that this na\"{i}ve instantiation constitutes the basis for a more advanced form of a privacy-anomaly detection system, analogous with \textit{k}-anonymity~\index{\textit{k}-anonymity} which constitutes the basis for more sophisticated formal privacy definitions.
The reasons are as follows. Firstly, this is an exploratory study to consider the question whether one can provide a privacy semantics for behavioural-based anomaly detection approaches or relate the notion of privacy-anomaly detection to the conventional definitions of privacy semantics?
Therefore, using a well-understood privacy model like \textit{k}-anonymity enables better understanding of the subject being explored and helps to avoid underlying complexities associated with other more complex privacy definitions.
Secondly, \textit{k}-anonymity served as a foundation of many subsequent formal privacy definitions, which is a good indicator of the applicability of this study onto other privacy definitions.

\subsection{A \textit{k}-Anonymity based Privacy-profile}

In the proposed model \textit{k}-anonymity is used to specify a privacy limit $[\![ \textit{k}$ , $q]\!]$,  whereby \textit{k} individual must share the same quasi identifier $q$ values in the result of a query.   
Intuitively, this means for that particular response, 
for a sufficient value of \textit{k}, an adversary can only narrow down to \textit{k} individuals. 
In the case where an adversary has a secondary dataset with overlapping quasi-identifier values, then the query response can be linked to \textit{k} different individuals, therefore minimizing the risk of re-identification.
In the model the privacy-profile is defined as a set of privacy limits. 
In terms of privacy, each privacy limit means that in a particular instance of a query response an adversary won't be able to distinguish an individual's quasi-identifier values from \textit{k} individuals for the set of quasi-identifiers that appeared in the query response.

\begin{table}
\centering
\begin{tabular}{| M{2.0cm}  | M{2.0cm} | M{2.0cm}| M{2.0cm}| M{2.0cm}|}
\hline
\texttt{age} & \texttt{zipcode} & \texttt{county} & \texttt{gender} & \texttt{salary}   \\ \hline \hline
\textgreater55  & 989234 & Cork & Male & 60K \\ \hline
\textgreater55  & 989234 & Cork & Male & 92K \\ \hline
\textgreater55  & 989234 & Cork & Male & 77K \\ \hline
\textgreater45  & 839523  & Cork & Male & 50K \\ \hline
\textgreater35 & 839777  & Dublin & Male & 60K \\ \hline
\textgreater35 & 839777  & Dublin & Male & 63K \\ \hline
\textgreater35 & 839777  & Dublin & Male & 85K \\ \hline
\textgreater35 & 839777  & Dublin & Male & 70K \\ \hline
\textgreater35 & 839777  & Dublin & Male & 60K \\ \hline
\textgreater50 & 839567  & Cork & Female & 72K \\ \hline
\textgreater50 & 839567  & Dublin & Female &  62K \\ \hline
\textgreater50 & 839567  & Cork & Female &  92K \\ \hline
\textgreater50 & 839567  & Dublin & Female &  77K \\ \hline
\textgreater50 & 839567  & Cork & Female &  68K \\ \hline
\end{tabular}
\caption{A fragment of relation \texttt{temp\_table}.}
\label{sourcetable}
\end{table}

Consider a relation \texttt{temp\_table}, as shown in Table~\ref{sourcetable}, having several attributes including a sensitive attribute \texttt{salary}, and quasi-identifiers \texttt{age}, \texttt{gender}, \texttt{zipcode}, and \texttt{county}. 
For ease of exposition we assume the values for attribute \texttt{age} are aggregated into age ranges, for instance, all the values for attribute \texttt{age} above $55$ are represented as \textgreater55.
Given a mined privacy limit $[\![ 3, \{\texttt{age}, \texttt{zipcode}\}]\!]$, in privacy-profile, then the response to the analyst query \texttt{SELECT age, zipcode FROM temp\_table WHERE gender = `Male' AND county = `Cork' AND age > 55;} as shown in Table~\ref{response:1} is not anomalous since the value of \textit{k} for the the quasi-identifiers \{\texttt{age}, \texttt{zipcode}\} in the response is greater than $3$. 
 
\begin{table}
\centering
\begin{tabular}{| M{3cm}  | M{3cm} | M{3cm}| }
\hline
\texttt{age} & \texttt{zipcode} & \texttt{salary}   \\ \hline \hline
\textgreater55  & 989234  & 60K \\ \hline
\textgreater55  & 989234  & 92K \\ \hline
\textgreater55  & 989234  & 77K \\ \hline
\end{tabular}
\caption{A relation $\mathscr{T}_{R1}$  resulting from the query \texttt{SELECT age, zipcode FROM temp\_table WHERE gender = `Male';}. }
\label{response:1}
\end{table}

\subsubsection{Mining \textit{k}-anonymity based Profiles for PAD} \label{p:3:1} ~\index{\textit{k}-anonymity)-PAD}
The privacy-anomaly detection consists of two phases, similar to traditional anomaly detection approaches, that are, learning phase and a detection phase. 
The instances of the privacy model are mined from audit logs in order to generate privacy-profiles. 
We refer to a privacy-profile that is mined from past logs in the learning phase as a normative privacy-profile.
The idea is to learn the \textit{k} values for sets of quasi-identifier(s) by mining past audit logs and interpret those mined `privacy limits' as `normal'.

Given an audit log $L^*$, consisting of query responses, $Pri(L^*)$ gives a privacy-profile consisting of privacy limits 
mined from log $L^*$, where $q$ $\in$ $QI$ represent a set of quasi-identifier.
A normative privacy-profile is generated from an anomaly-free past log $L^*_{norm}$ and is denoted by 
$Pri(L^*_{norm})$ = \{ $[\![ \textit{k}_1 , q_1 ]\!]$, $[\![ \textit{k}_2 , q_2 ]\!]$, \dots, $[\![ \textit{k}_m , q_m ]\!]$ \}.
For example, consider the relation $\mathscr{T}_{R2}$ shown in Table~\ref{resulttable}, the mined value of \textit{k}  for the set of quasi-identifiers \{\texttt{age, zipcode, county}\} is $4$, 
that is, $[\![4, \{age, zipcode, county\}]\!]$ 
$\in$ $Pri(L^*_{norm})$. 
In essence we are constructing privacy limit $(L^*,q)$ which returns \textit{k} as a limit to the privacy in the table for a given $q$.
The normative privacy-profile is effectively a set of these privacy limits mined against the logs for a given set of quasi-identifiers. 
Intuitively, the tuples in the normative privacy-profile shows to what extent one narrows down to individuals records in normative settings.

\begin{table}
\centering
\begin{tabular}{| M{2cm}  | M{2cm} | M{2cm}|  M{2cm}| }
\hline
\texttt{age}        & \texttt{zipcode} & \texttt{county} & \texttt{salary}   \\ \hline \hline
\textgreater55    & 839523  & Cork & 60K \\ \hline
\textgreater55    & 839523  & Cork & 92K \\ \hline
\textgreater55    & 839523  & Cork & 77K \\ \hline
\textgreater45    & 839523  & Cork & 50K \\ \hline
\textgreater35 & 839777  & Dublin & 60K \\ \hline
\textgreater35 & 839777  & Dublin & 63K \\ \hline
\textgreater35 & 839777  & Dublin & 85K \\ \hline
\textgreater35 & 839777  & Dublin & 70K \\ \hline
\textgreater35 & 839777  & Dublin & 60K \\ \hline
\end{tabular}
\caption{A relation $\mathscr{T}_{R2}$  resulting from the query \texttt{SELECT age, zipcode, county FROM temp\_table WHERE gender = `male';}. }

\label{resulttable}
\end{table}

\subsubsection{Detecting Privacy-anomalies}
The detection phase, in terms of privacy, checks if an adversary is able to narrow down to fewer than \textit{k} individuals for a given set of quasi-identifiers in the normative profile. 
In the instance, where the adversary is able to narrow down to fewer than specified \textit{k} individuals for a given set of quasi-identifier then this instance is labelled as a privacy-anomaly and poses higher risk of re-identification relative to normal. 
During the detection phase, the run-time profile $Pri(L^*_{run})$ constructed given a run-time log $L^*_{run}$. 
$Pri(L^*_{run})$ is the constructed run-time profile. 
Given privacy limits $[\![k_i, q_i]\!]$ and $[\![k_j, q_j]\!]$ then $[\![k_i, q_i]\!]$ subsumes $[\![k_j, q_j]\!]$ (denoted $[\![k_i, q_i]\!]$ $\preceq$  $[\![k_j, q_j]\!]$) if imposing privacy limit $[\![k_j, q_j]\!]$ instead of $[\![k_i, q_i]\!]$ leads to no additional loss of privacy. 
Formally, \[ [\![k_i, q_i]\!] \preceq  [\![k_j, q_j]\!] \equiv q_i \subseteq q_j \land k_j \geq k_i \] 

In the case where  $[\![k_i, q_i]\!]$ $\in$ $Pri(L^*_{norm})$ and $[\![k_j, q_j]\!]$ $\in$ $Pri(L^*_{run})$ then $[\![k_i, q_i]\!]$ $\preceq $ $[\![k_j, q_j]\!]$ means that $[\![k_j, q_j]\!]$ can be safely replaced by $[\![k_i, q_i]\!]$ without any loss of privacy.
If a privacy limit subsumes another intuitively it means if the subsumed privacy limit is replaced by the one that subsumes it then there is no loss of privacy. 

Consider the response of a query at run-time shown in Table~\ref{resulttable2}, and that there exists a privacy limit $[\![3, \{age, zipcode\}]\!]$ in $Pri(L^*_{norm})$. 
The mined value \textit{k} of the set of quasi-identifier $\{age, zipcode\} $ is greater than $3$ therefore this privacy limit $[\![5, \{age, zipcode\}]\!]$ in $Pri(L^*_{run})$ is considered to be subsumed by the privacy limit $[\![3, \{age, zipcode\}]\!]$ in $Pri(L^*_{norm})$. 
In terms of privacy, it means given that this instance of query response an adversary can narrow down so many individuals as one normally is able to for a given set of quasi-identifiers.

\begin{table}
\centering
\begin{tabular}{| M{3cm}  | M{3cm} | M{3cm}| }
\hline
\texttt{age}            & \texttt{zipcode} & \texttt{salary}   \\ \hline \hline
\textgreater50 & 839567  & 72K \\ \hline
\textgreater50 & 839567  & 62K \\ \hline
\textgreater50 & 839567  & 92K \\ \hline
\textgreater50 & 839567  & 77K \\ \hline
\textgreater50 & 839567  & 68K \\ \hline
\end{tabular}
\caption{A relation $\mathscr{T}_{R3}$ resulting from the query \texttt{SELECT age, zipcode FROM temp\_table WHERE gender = `female';}. }

\label{resulttable2}
\end{table}

\section{Security-anomaly Detection System Detecting Privacy-anomalies} \label{pva}
This section explores whether privacy-anomalies (as identified by the model in Section~\ref{p:3:1}) are also identified as security-anomalies by a security-anomaly detection system in~\cite{10}).
The security-anomaly detection system in~\cite{10} relies on n-grams to construct profiles of querying behaviours using audit logs of SQL queries. 
The system in~\cite{10} effectively detects malicious accesses by insider to a database management system.
We consider a variation of the hospital dataset, a fragment of the dataset is shown in Table~\ref{sourcetabelexp}. 
Logs were generated for construction of a normative profile and another for the construction of a run-time profile.
The training logs (anomaly-free) for the n-gram based approach are denoted by 
$L^{hosp}_{norm}$, while the anomalous run-time logs for the hospital datasets are denoted by 
$L^{hosp}_{run}$. 
The next section studies whether a security-anomaly detection system detects privacy-anomalies identified by the privacy-anomaly detection system for these dataset. 

\begin{figure}[h]
  \includegraphics[width=\linewidth]{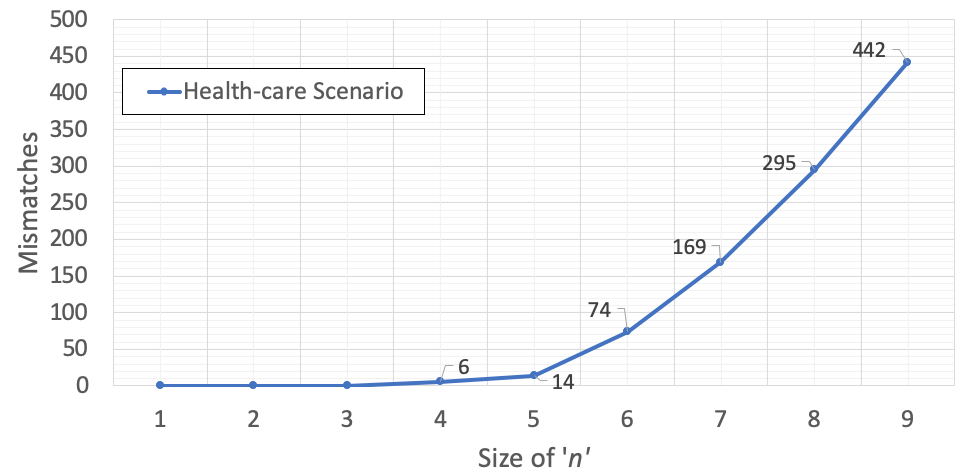}
  \caption{The figure shows the number of mismatches between $ngram(L^{hosp}_{test1}, n)$ and $ngram(L^{hosp}_{test2}, n)$ for different values of \textit{n}.}
  \label{tune2}
\end{figure}

To construct normative and run-time profiles using the n-gram model, selection of an appropriate value of the size of n-gram was desirable for the hospital dataset.
To select an appropriate size of an n-gram in this scenario, test logs $L^{hosp}_{test1}$ and $L^{hosp}_{test2}$ were generated in a safe environment (anomaly-free). 
N-gram profiles were constructed with varying n-gram size, that are, $ngram(L^{hosp}_{test1}, n)$ and $ngram(L^{hosp}_{test2}, n)$, and generated profiles were compared. 
Figure~\ref{tune2} depicts the number of n-gram mismatches arising when comparing the normal test $ngram(L^{hosp}_{test1}, n)$ and 
$ngram(L^{hosp}_{test2}, n)$, for different values of n.
From the experiments, the n-gram of the size of 4 (\textit{n} = 4) was considered optimal as it resulted in an acceptable number of mismatches.

Once the value of \textit{n} was decided upon, the normative and run-time profiles were constructed for the experiments. 
Given the training logs $L^{hosp}_{norm}$ and $L^{hosp}_{run}$ n-gram profiles were constructed such that $ngram(L^{hosp}_{norm}, 4)$ and $ngram(L^{hosp}_{run}, 4)$, and subsequently the normative and runtime profiles were compared. 

The same queries in logs $L^{hosp}_{norm}$ and $L^{hosp}_{run}$ were executed in the presence of the privacy-anomaly detection system (described in Section~\ref{pad}) resulting in logs of query responses $L^{hosp*}_{norm}$ and $L^{hosp*}_{run}$. 
Subsequently, a normative privacy-profile $Pri(L^{hosp*}_{norm})$ and a run-time $Pri(L^{hosp*}_{run})$ profiles were constructed and compared.

The attribute \texttt{patient\_ID} and \texttt{e-mail\_ID} were considered as a unique identifier, the attribute \texttt{diagnosis} was considered as a sensitive attribute while the rest of the attributes including \texttt{first\_name}, \texttt{last\_name}, \texttt{status}, \texttt{dob}, \texttt{gender}, \texttt{city}, and \texttt{marital\_status} were considered as quasi-identifiers. 
For the experimentation, two categories of privacy-anomalies were injected as described in Table~\ref{tableanomalies}. 
Using this anomaly-containing run-time log, from $15$ privacy-anomalies $13$ were detected by the n-gram based security-anomaly detection system proposed in~\cite{10} and the  
privacy-anomaly detection system proposed in this paper.

\begin{table}
\centering
\begin{tabular}{| M{1.8cm}  | M{1.8cm} | M{1.8cm}|  M{1.8cm}: M{1.8cm}: M{1.0cm}|} 
\hline
\texttt{dob} & \texttt{city} & \texttt{gender}  & \texttt{diagnoses} & \texttt{country}&... \\ \hline \hline
1981    & Dublin  & Male & Flu & Ireland&... \\ \hline
1981    & Dublin  & Male & Flu & Ireland&... \\ \hline
1981    & Dublin  & Male & Diarrhoea & Germany&... \\ \hline
1920    & Cork    & Male & Heart Disease & Ireland&... \\ \hline
1981    & Galway    & Female & Acne & Ireland&... \\ \hline
1984    & Galway    & Male & Flu & Spain&... \\ \hline
1984    & Galway    & Male & Diabetes & Ireland&... \\ \hline
1984    & Galway    & Male & Hypertension & Ireland&... \\ \hline
1984    & Galway    & Male & Leg Fracture & France&... \\ \hline
...    & ...  & ... & ... & ...&... \\ \hline
...    & ...  & ... & ... & ...&... \\ \hline
\st{1981}   & \st{Dublin}  & \st{Male} & \st{Flu} & \st{Germany}&... \\ \hline
\end{tabular}
\caption{A fragment of hospital dataset.The strike-through attribute values represents a deleted row.}
\label{sourcetabelexp}
\end{table}

\begin{table}
\centering
\begin{tabular}{| l  | c| }
\hline
Description of privacy-anomalies  &   \specialcell{Number of \\ anomalies injected}\\ \hline \hline
\specials{Addition of one or more attributes to the base relation 
\\ shown in Table~\ref{sourcetabelexp}.  For instance, a new attribute, 
\\ like \texttt{country}, was inserted in the relation and queries 
\\ were made to retrieve this attribute values.}  &  $5$  \\ \hline
\specials{Update or Deletion of records from relation 
\\ shown in Table~\ref{sourcetabelexp} } &  $10$  \\ \hline
\end{tabular}
\caption{Description of Privacy-anomalies injected.}

\label{tableanomalies}
\end{table}

\subsection{Detected Privacy-anomalies}
The n-gram based security-anomaly detection system detected all those privacy-anomalies that were generated by injecting one more attribute into the relation. 
The privacy-anomalies injected by adding one more attribute were identified as privacy-anomalies by both systems. 
The reason that they were identified was because there were no n-gram 
that contained a reference to new attribute in its query abstraction. 

One of the detected privacy-anomalies 
corresponds to the query shown below.
\\
\texttt{SELECT diagnoses, dob, city, country} \\
\texttt{FROM hospitalDB} \\
\texttt{WHERE dob = ‘1981’ }\\
\texttt{     AND city = ‘Dublin’};\\

The normative privacy-profile contains no privacy limit reference to the new (or combination of new) attribute.

\subsection{Undetected Privacy-anomalies}
A privacy-anomaly undetected by the n-gram based approach but detected by the privacy model is:
\\
\texttt{SELECT dob, city, diagnoses} \\
\texttt{FROM hospitalDB} \\
\texttt{WHERE dob = ‘1920’}\\
\texttt{     AND city = ‘Cork’ ;}\\

\begin{table}
\centering
\begin{tabular}{| M{3cm} | M{3cm} | M{3cm} | }
\hline
\texttt{dob}  &\texttt{city}&\texttt{diagnoses}       \\ \hline \hline
1920&Cork&Heart Disease \\ \hline
\end{tabular}
\caption{Response to a undetected privacy-anomalous query.}
\label{resundetect}
\end{table} 

The query returns a relation with one record as shown in Table~\ref{resundetect}.  
It is identified as a privacy-anomaly by the privacy model for the reason being that the specified value of \textit{k} for the specified set of quasi-identifier  
meant that an adversary was able to single out an individual.
This anomaly is undetected by n-gram based security-anomaly detection approach because there was an n-gram in normative profile contained a reference to this
query abstraction.

\subsection{Identifying Appropriate Privacy Limits}
In order to find the optimal values of \textit{k}, in the mining process, in theory, all the combinations of quasi-identifiers need to be considered. 
This, in essence, is a combinatorial explosion, especially in the case of a large number of quasi-identifiers.
Additionally, one may discover either very large or very small values of \textit{k} in practice for certain combinations of quasi-identifiers. 
Therefore, in order to discover reasonable values of \textit{k}, one may define a range while mining the values of \textit{k} such that the values falling within the range and their corresponding combinations of quasi-identifiers are considered for privacy-profiles.

\section{Privacy Attacks based on Query Sequences } \label{bpc}
This section  demonstrates a privacy attack whereby an adversary discovers information about an individual while the privacy-anomaly detection system is in place. 
Consider the relation shown in Table~\ref{ex5}. 

\begin{table*}[h]\arrayrulecolor{white}
\centering
       \begin{tabular}{|l|l|l|l|l|}
        \hline
        \color{white}
      \cellcolor{black}  \texttt{Name} &  \cellcolor{black}  \texttt{MaritalStatus} & \cellcolor{black} \color{white} \texttt{City} & \cellcolor{black}\color{white} \texttt{Age} &\cellcolor{black} \color{white} \texttt{\specials{Salary \\ (sensitive \\ attribute)}}   \\ \arrayrulecolor{white} \hline \hline
        \rowcolor{babyblueeyes} 
       \texttt{Mark} & \texttt{Single}    & \texttt{New York} &\texttt{[30 - 40]}&\texttt{112k}  \\ \arrayrulecolor{white} \hline
\rowcolor{babyblueeyes}      
     \texttt{James} &  \texttt{Single}    & \texttt{New York } &\texttt{[30 - 40]}&\texttt{34k}  \\ \hline
\rowcolor{babyblueeyes}     
    \texttt{John} &  \texttt{Single}  & \texttt{New York} &\texttt{[30 - 40]}&\texttt{56k} \\ \hline
\rowcolor{babyblueeyes}     
\texttt{Henry} &         \texttt{Single}   & \texttt{New York} &\texttt{[30 - 40]}&\texttt{78k} \\ \hline
\rowcolor{babyblueeyes}     
  \texttt{Imran} &    \texttt{Single}   & \texttt{New York} &\texttt{[30 - 40]}&\texttt{91k}\\  \hline
      \rowcolor{babyblueeyes}  
      \texttt{David} & \texttt{Married}   & \texttt{London} &\texttt{[30 - 40]}&\texttt{112k}\\  \hline
        \rowcolor{babyblueeyes}  
    \texttt{Alice} &    \texttt{Married}   & \texttt{London} &\texttt{[30 - 40]}&\texttt{30k}\\  \hline
     \rowcolor{babyblueeyes}  
   \texttt{Bob} &    \texttt{Married}   & \texttt{London} &\texttt{[30 - 40]}&\texttt{45k}\\  \hline
     \rowcolor{babyblueeyes}  
    \texttt{Aron} &    \texttt{Married}   & \texttt{London}  &\texttt{[30 - 40]} &\texttt{115k}\\  \hline
       \rowcolor{babyblueeyes}  
      \texttt{Harry} &  \texttt{Married}   & \texttt{London} &\texttt{[30 - 40]} &\texttt{180k}\\  \hline
          \rowcolor{babyblueeyes}  
      \texttt{Jordan} &  \texttt{Separated}   & \texttt{Cork} &\texttt{>40} &\texttt{65k}\\  \hline
          \rowcolor{babyblueeyes}  
  \texttt{Ryan} &      \texttt{Separated}   & \texttt{Cork} &\texttt{>40} &\texttt{100k}\\  \hline
        \rowcolor{babyblueeyes}  
\texttt{Bentley} &        \texttt{Separated}   & \texttt{Cork} &\texttt{>40} &\texttt{80k}\\  \hline
          \rowcolor{babyblueeyes}  
             \texttt{Simon}  & \texttt{Married} & \texttt{Rennes}  & \texttt{>40} & \texttt{150k}\\  \hline
    \end{tabular}
\caption{Relation \texttt{updated\_table\_smp}.}
    \label{ex5}
\end{table*} 

\begin{table*}[h] \arrayrulecolor{black}
\centering
\begin{tabular}{|c|l|}
\hline
 $Q_1$  & \specials{\texttt{SELECT}  \texttt{Salary} \\  \texttt{FROM}  \texttt{updated\_table\_smp}  \\ \texttt{WHERE}  \texttt{city} =  \texttt{‘Rennes’};} \\ \hline

 $Q_2$   &  \specials{\texttt{SELECT}  \texttt{Salary} \\  \texttt{FROM}  \texttt{updated\_table\_smp}; }  \\ \hline

 $Q_3$  & \specials{ \texttt{SELECT}  \texttt{MaritalStatus},  \texttt{Salary},  \texttt{Age}   \\ \texttt{FROM}  \texttt{updated\_table\_smp}  \\ \texttt{WHERE}  \texttt{City}= \texttt{‘New York’} \texttt{AND} \texttt{city} =\texttt{‘London’}  \texttt{AND}  \texttt{city} = \texttt{‘Cork’;}} \\ \hline

\end{tabular}
\caption{Sequence of queries executed over the relation \texttt{updated\_table\_smp} shown in Table~\ref{ex5}. }
\label{query2smp}
\end{table*}

Suppose the sequence of queries $Q_1$, $Q_2$ and $Q_3$, shown in Table~\ref{query2smp}, are executed over the relation  \texttt{updated\_table\_smp} shown in Table~\ref{ex5}.

Query $Q_1$ is labelled as an privacy-anomaly (and the query response is suppressed) because the threshold is not satisfied 
as \textit{k}-anonymity and $\mbox{\textit{DRSQL}}$ is not satisfied by the query $Q_1$.
Whereas query $Q_2$ passes the threshold and the response to the query is shown in Table~\ref{ex6} that results in $14$ records, that is, the value of attribute \texttt{Salary}, being returned.
Query $Q_3$ also passes the privacy criteria and returns $13$ records, as shown in Table~\ref{ex7}. 
The adversary, knowing that \texttt{Simon's} record is in the table (as background/external knowledge) and that Simon lives in Rennes, reveals that last remaining entry blocked by the query mechanism is of `\texttt{Simon}' and the corresponding salary attribute value is $150$k.

In particular, the described attack is a form of a differencing attack~\cite{279}. 
Differencing attacks have been seen previously on aggregates. 
This demonstrates that \textit{k}-anonymity in interactive settings is also susceptible to these differencing attacks. 
For the purpose of demonstration, the example of a differencing attack is kept simple; however, real world differencing attacks can take more sophisticated forms, where the adversary can make multiple queries to narrow down the aggregate data until the subject's information is not revealed.

\begin{table}[h]\arrayrulecolor{black}
\centering
    \begin{tabular}{|c|}
        \hline \color{black}
     \cellcolor{white} 
         \texttt{Salary} \texttt{(sensitive attribute)}   \\  \hline \hline 
 \texttt{112k} \\ \hline
\texttt{34k} \\  \hline
       \texttt{56k} \\ \hline
    \texttt{78k} \\ \hline
    \texttt{91k}\\  \hline
      \texttt{112k}\\  \hline
        \texttt{30k}\\  \hline
        \texttt{45k}\\  \hline
     \texttt{115k}\\  \hline
       \texttt{65k}\\  \hline
         \texttt{100k}\\  \hline
        \texttt{80k}\\  \hline
         \texttt{150k}\\  \hline
    \end{tabular}
\caption{Records returned in the response to query $Q_2$.}
    \label{ex6}
\end{table}

\begin{table*}[ht]\arrayrulecolor{black}
\centering
    \begin{tabular}{|c | c | c | c| }
        \hline
          \texttt{   MaritalStatus    } & \texttt{     City      } &  \texttt{     Age    } & \texttt{\specials{Salary (sensitive attribute)}}   \\  \hline \hline
       \texttt{Single}    & \texttt{New York}   & \texttt{[30 - 40]}  & \texttt{112k}  \\ \hline
       \texttt{Single}    &\texttt{New York}   & \texttt{[30 - 40]}  & \texttt{34k}  \\ \hline
       \texttt{Single}    & \texttt{New York}  & \texttt{[30 - 40]}  & \texttt{56k} \\ \hline
       \texttt{Single}    & \texttt{New York}  & \texttt{[30 - 40]}  & \texttt{78k} \\ \hline
      \texttt{Single}     & \texttt{New York}   & \texttt{[30 - 40]}  & \texttt{91k}\\  \hline
      \texttt{Married}  & \texttt{London}   & \texttt{[30 - 40]} & \texttt{112k}\\  \hline
      \texttt{Married}  & \texttt{London}   & \texttt{[30 - 40]}  & \texttt{30k}\\  \hline
      \texttt{Married}  & \texttt{London}   & \texttt{[30 - 40]} & \texttt{45k}\\  \hline
      \texttt{Married}  & \texttt{London}   & \texttt{[30 - 40]} & \texttt{115k}\\  \hline
      \texttt{Married}  & \texttt{London}   & \texttt{[30 - 40]} & \texttt{180k}\\  \hline
        \texttt{Separated}   & \texttt{Cork}  & \texttt{>40} & \texttt{65k}\\  \hline
        \texttt{Separated}   & \texttt{Cork}  & \texttt{>40} & \texttt{100k}\\  \hline
        \texttt{Separated}   & \texttt{Cork}  & \texttt{>40} & \texttt{80k}\\  \hline
    \end{tabular}
\caption {Records returned in response of query $Q_3$.}
    \label{ex7}
\end{table*}

\section{Applying Security-anomaly Detection to Detect Unknown Privacy Attacks} \label{7.7}
In general, interactive query mechanisms are susceptible to the attacks described in the previous section, and as a consequence there is little privacy-preserving interactive query mechanisms (specifically for microdata release) in the existing literature.
The existing differentially private interactive mechanisms allow a limited number of interactive queries for aggregate data. 
Restricting the number of queries is a significant barrier for an analyst in achieving the true potential for data analytics. 
Privacy attacks, similar to the one presented in the previous section, are violations of formal privacy definitions like \textit{k}-anonymity.

Another aspect of these privacy attacks is that the querying pattern to infer information about the subject(s) is unknown, therefore, we refer to them as unknown privacy attacks.
Unknown privacy attacks lead to inferring information about the subject(s). 
An adversary can articulate the queries in different ways to reveal information about a subject(s).
In this work, inference implies privacy attacks, that is, the adversary infers information about the subject(s) while a formal privacy definition is in place resulting in a violation of formal privacy definition. 
Additionally, these privacy attacks are based on query correlation, that is, an individual query is safe in terms of privacy but when considered as sequence they result in the violation of formal privacy definition. 
This section presents an investigation into whether the inferences can be detected as anomalies. 

We present a novel perspective on the detection of privacy attacks by proposing an interpretation of the behavioural-based detection approach. 
There are a number of behavioural-based anomaly detection detection approaches that can be explored in this context~\cite{10, 235, 288, 64}.
We investigate the application of n-gram approach, proposed approach in~\cite{10}, to detect unknown privacy attacks as anomalies in the next section.

\subsection{Detecting Privacy Attacks as Privacy-Anomalies}
A behavioural-based approach to detect inferences as anomalies is described in this section where the n-gram based approach is applied to the audit logs of the (\textit{k}-anonymity)-PAD system. 
The idea is to model querying behaviours in the presence of a privacy-preserving interactive query mechanism and compare the normative querying behaviour with the run-time querying behaviour to detect deviations.

For the SQL query abstraction, the specialization of the abstraction, as discussed in ~\cite{10}, is employed. 
The SQL query abstraction that is a tuple representation of an SQL query and consists of query features like relation name, attribute names, the amount of returned data or any statistics on the returned data. 

An abstraction of an SQL query $Q_i$ is denoted as $Abs(Q_i)$.~\index{SQL query abstraction}
The adopted query abstraction technique has also been studied in~\cite{17}.
The query abstraction technique replaces the constant values in a query $Q_i$ with place-holders (literal `VAR\_VAL'), and is denoted as $Abs(Q_i)$. 
$Abs(L)$ is defined as the mapping of $Abs(Q_i)$ over the elements $Q_i$ of $L$. 
The reason to chose this query abstraction is that it gives us a reasonable level of precision in capturing the querying behaviour of user.
A more fine grained abstraction would require some symbolic evaluation of the queries which was beyond the scope of this work.
Examples of the employed SQL query abstraction technique are shown in Table~\ref{qmaps}. 

\begin{table*} \arrayrulecolor{black}
\centering
\begin{tabular}{|c|l|l |  } \hline 
   \color{black}
$Q_i$ & SQL statement &  SQL query abstraction $Abs(Q_i)$
\\ \hline \hline
$Q_1$ & \texttt{\specials{SELECT city \\ FROM bankDatabase \\ WHERE id = 2}}  & \texttt{\specials{SELECT city \\ FROM bankDatabase \\ WHERE id = VAR\_VAL}}
\\ \hline
$Q_2$ & \texttt{\specials{SELECT city \\ FROM bankDatabase  \\ WHERE id = 9}}  & \texttt{\specials{SELECT city \\ FROM bankDatabase \\ WHERE id = VAR\_VAL}} 
\\ \hline
$Q_3$ & \texttt{\specials{ SELECT city \\ FROM bankDatabase  \\ WHERE id = 3} } & \texttt{\specials{SELECT city \\ FROM bankDatabase \\ WHERE id = VAR\_VAL} }
\\ \hline
$Q_4$ & \texttt{\specials{SELECT city \\ FROM bankDatabase \\ WHERE id =  3 AND \\ Name = "Alice"}  	}		& \texttt{\specials{SELECT city \\ FROM bankDatabase \\ WHERE id = VAR\_VAL  AND \\ Name = VAR\_VAL}}			\\ \hline
\end{tabular}
\caption{Examples of deployed SQL abstractions.} \label{qmaps}
\end{table*}
 
The n-gram profiles are generated in the same manner as discussed in~\cite{10} that is given a safe audit log of SQL query $L^{pp}_{norm}$ and a run-time log $L^{pp}_{run}$ then the constructed normative profile and run-time profile are $\beta_{norm}$ = $ngram(Abs(L^{pp}_{norm}), n)$ and $\beta_{run}$ = $ngram(Abs(L^{pp}_{run}), n)$.
The mismatches are given by $S^{miss}_{\beta_{run} - \beta_{norm}}$ = $\beta_{run}$ - $\beta_{norm}$.

In order to evaluate the detection of privacy attacks by applying the n-gram based approach to the logs of interactive querying mechanism, a synthetic query generator was designed that had defined a set of SQL query templates. 
The underlying database was populated with a fragment version of well-known Census (Adult) dataset~\cite{192}.
Query templates were designed to be executed on the Census dataset.
The queries were count queries mimicking a data analytics scenario. 
For example, the count queries were for the form: how many subjects are Female, how many subjects have a Bachelors degree, so on and so forth.
For the experimentation, a safe log $L^{pp}_{norm}$ was generated for the construction of normative profile using the synthetic data generator.

In order to construct privacy attacks, five unique records were inserted into the database that leads to inferences, where unique implies that one of the attribute or combination of the attributes existed only once in the entire database.
For example, a record was inserted with \texttt{occupation} as \texttt{post-doc}, that is, in the underlying database there was only one record where the value for \texttt{occupation} was \texttt{post-doc}.
Another record was inserted where the value for \texttt{native-country} was set to \texttt{Malaysia}, that is, there was only one record where the \texttt{native-country} was \texttt{Malaysia}.
Table~\ref{infer:privattackm} shows the make-up of the inserted records to enable privacy attacks. 

\begin{table}[h]
\centering
\begin{tabular}{|c|l|} 
\hline
 \# & Description of Unique Record \\ \hline \hline
 1&  Attribute \texttt{occupation} with value as \texttt{post-doc} \\ \hline
 2& Attribute \texttt{native-country} with value as \texttt{Malaysia} \\ \hline
 3& Attribute \texttt{native-country} with value as \texttt{Spain} and Attribute \texttt{age} as \texttt{33} \\ \hline
 4& \specials{Attribute \texttt{native-country} with value as \texttt{Singapore} \\ and Attribute \texttt{age} as \texttt{32} } \\ \hline
 5& \specials{Attribute \texttt{native-country} with value as \texttt{Singapore} \\ and Attribute \texttt{occupation} as \texttt{Academics}} \\ \hline
 \end{tabular}
 \caption{Inserted unique records in the database to enable privacy attacks.}
  \label{infer:privattackm}
\end{table}

Queries were made to infer the associated salaries for these five unique records.
Table~\ref{infer:privattack} shows the number of queries made to reveal the salary for the records shown in Table~\ref{infer:privattackm}.
These malicious query sequences were made part of the other logs $L^{pp}_{run}$ for the construction for the run-time profile.

\begin{table}[h]
\centering
\begin{tabular}{|c|c|c|}
\hline
Privacy Attack \# & Inference Sequence Length  \\ \hline \hline
1&7 \\ \hline
2&9 \\ \hline
3&17 \\ \hline
4&13 \\ \hline
5&10 \\ \hline
\end{tabular}
\caption{Length of the query sequences to reveal salaries.}
\label{infer:privattack}
\end{table}

A normative profile $\beta_{norm}$ and a run-time profile $\beta_{run}$ were constructed using $L^{pp}_{norm}$ and $L^{pp}_{run}$, respectively and compared.
The size of the n-gram was kept at $4$ for the generation of profiles. 
The sequence of queries made to infer injected unique records was labelled as anomalies the detection phase by the n-gram approach. 
The Table ~\ref{infer:privattackdetect} shows the number of mismatches for each privacy attack.

\begin{table}[h]
\centering
\begin{tabular}{|c|c|}
\hline
Privacy Attack \# & |$S^{miss}_{\beta_{run} - \beta_{norm}}$|  \\ \hline \hline
1 & 5 \\ \hline
2 & 7 \\ \hline
3 & 19 \\ \hline
4 & 15 \\ \hline
5 & 13 \\ \hline
\end{tabular}
\caption{Detection of privacy attacks as privacy-anomalies: the table shows the number of mismatches that resulted from each of the 5 privacy attacks with the n-gram of size $4$.}
\label{infer:privattackdetect}
\end{table}

The query sequences resulting in inferences were detected as privacy-anomalies, which is a indication of potential effectiveness of n-gram based approach to detect inference and unknown privacy attacks as privacy-anomalies.

\section{Conclusions} \label{p:c}
While existing behavioural-based anomaly detection systems considered anomalies arising from anomalous queries of users, this paper explores anomalies characterised in terms of formal definitions of privacy. 
This work studies privacy semantic notion of behavioural-based anomaly detection systems. 
The notion of privacy-anomaly detection (PAD) introduced in this work enables one to learn a privacy model from the past log of interaction with the DBMS and detects deviations as privacy-anomalies.
A na\"ive instantiations of PAD was presented based on \textit{k}-anonymity (\textit{k}-anonymity, DR)-PAD. 
A study was carried out to examine whether the privacy violations based on a single query detected the privacy-anomaly detection system are also detected by n-gram security-anomaly detection system as anomalies. 
Results showed a number of single query based privacy violations that had no reference n-gram in normative profiles were labelled as anomalies by n-gram based anomaly detection system. 
This work also considered privacy attacks that were violations of formal privacy definitions and were based on query correlation where a single query is not privacy-anomalous but a sequence of queries results in a violation of formal privacy definition.
Results showed that behavioural-based security anomaly detection system (constructed using n-grams) in~\cite{10} detected these privacy attacks as privacy-anomalies.  
This led to a discovery of another aspect of the n-gram based approach whereby when it is applied on the logs generated by interactive query settings with the presence of formal privacy definition, it has the potential to detect privacy attacks based on query correlation as privacy-anomalies.

Privacy attacks can manifest itself in a variety of unexpected ways, the results suggest that therefore a silver bullet for anonymization may not be a way forward rather utilise defence in depth from privacy perspective. 
One such privacy control is the proposed privacy anomaly detection system.  
As a topic of future work, we plan to explore how to compose and compare multiple privacy definitions using multi-criteria decision-making method found in fuzzy logic~\cite{203,204}, in particular, known as \textit{triangular-norms} (\textit{t-norms}).

\bibliographystyle{unsrt}
\bibliography{Imran_Submission} 

\end{document}